\documentclass[10pt,times,twoside,twocolumn,epsf]{article} 
\usepackage{times,fullpage}
\usepackage{epsfig,url}
\usepackage[bf]{subfigure}

\begin{document}

\title{Non-uniform Information Dissemination for Sensor
Networks\thanks{An earlier version of this paper appeared in the
International Conference on Networking Protocols (ICNP 2003)~\cite{tilak-icnp03}. Significant changes in this paper include the addition of an analysis of the protocol in the presence of mobility, as well as an expanded discussion section.}}

\author{Sameer Tilak$^\dag$, Amy Murphy$^\ddag$, Wendi Heinzelman$^\ddag$ and  Nael B. Abu-Ghazaleh$^\dag$ \\
\  \\
\begin{minipage}{80mm}
\begin{center}
        $^\dag$Computer System Research Laboratory \\
        Dept. of CS, Binghamton University \\
        Binghamton, NY~~13902--6000 \\
    \url{{sameer, nael}@cs.binghamton.edu}
\end{center}
\end{minipage}
\begin{minipage}{80mm}
\begin{center}
    $^\ddag$Computer Science Department \\
    University of Rochester  \\
    Rochester, NY~~14627--0126 \\
    \url{{murphy,wheinzel}@cs.rochester.edu}
\end{center}
\end{minipage}
}

\maketitle

\begin{abstract}

Future smart environments will be characterized by multiple nodes that
sense, collect, and disseminate information about environmental phenomena
through a wireless network.  In this paper, we define a set of applications
that require a new form of distributed knowledge about the environment,
referred to as \textit{non-uniform information granularity}.  By non-uniform
information granularity we mean that the required accuracy or precision of
information is proportional to the distance between a source node (information
producer) and current sink node (information consumer). That is, as the
distance between the source node and sink node increases, loss in information
precision is acceptable.  Applications that can benefit from this type of
knowledge range from battlefield scenarios to rescue operations.  The main
objectives of this paper are two-fold: first, we will precisely define
non-uniform information granularity, and second, we will describe 
different protocols that achieve non-uniform information dissemination and
analyze these protocols based on complexity, energy consumption, and
accuracy of information.
\end{abstract}

\section{Introduction}

Smart environments are characterized by a large number of distributed
sensors that collect environmental information 
and disseminate that information across wireless links.  Typically,
sensor networks focus on collecting this information in a single place
for analysis, taking into consideration optimizations from local
aggregation (e.g., LEACH~\cite{heinzelman-99}) or processing data en
route to a central location (e.g., MagnetOS~\cite{magnetos}).  While
such central collection is important for many applications, it does
not match the requirements of all applications that can exploit sensor
networks.

For example, consider a military application with sensors distributed
throughout an area collecting information about passing vehicles, air
contaminants, land mines, and other environmental data.  We assume the
sensors can communicate with one another, and a soldier that moves
throughout the region can contact any nearby sensor to find out both
the state of that sensor, as well as any other information it has
collected from the other networked sensors. For this soldier, clearly
the events occurring in the immediate neighborhood are most important.
For example, it is more critical to know about a land mine nearby than
one several miles away.  Nonetheless, it is still important that the
soldier have a general overview of the area in order to plan and make
appropriate decisions.  Similarly, consider a rescue scenario where a
team of fire fighters is working to rescue trapped victims.  In this
case, the fire fighters require precise information about their
immediate surroundings in order to make decisions about using
resources to make progress, as well as some global knowledge to plan
a path to the victims as well as an escape path back to safety. In
the above applications, sensors are static and mobile users 
connect to the nearby sensors to obtain the required information.
However, one can imagine a network where mobile users themselves are
carrying sensors placed on them. In this paper, we focus on protocols for
static sensor networks (sensors are stationary), but then show that
the protocols are resilient to mobility.

The applications above differ from those typically studied for sensor network
applications in the following way: the information is not collected
centrally, but instead it is utilized at several places in the network
(e.g., the locations of the individuals).  While some sensor network
applications accomplish this in a query driven manner, asking a
central source for the latest collected information, these
applications require continuous updates.  A simplistic solution to
this problem is to proactively flood updates from each sensor to every
other sensor.  This solution is extremely inefficient and does not scale to
large numbers of sensors.
In the scenarios we target, 
information from a particular sensor is most important 
to those surrounding it, with the value of the information decreasing as a
function of distance from the sensor.  Specifically, the necessary precision
of information is proportional to the distance between an information
producer and an information consumer.  We refer to such a requirement
as a \textit{non-uniform information dissemination} requirement, a new
concept we introduce here.

This paper introduces and analyzes several protocols that perform
non-uniform information dissemination.  As these protocols are
intended to run on wireless sensor networks, they must abide by the
requirements of that environment, namely they must be energy efficient and 
have low complexity.  The distinguishing feature of this new
application class is that it is possible to trade accuracy of disseminated
information for energy.  Our experimental results clearly show this
trade-off using a number of different protocols.

The remainder of this paper begins with a characterization of the
requirements for non-uniform information dissemination protocols.
Section~\ref{protocols} describes the details of several protocols.
In section~\ref{experimentation}, we discuss the implementation
details of the protocols within the ns-2 simulator and then present
our experimental results, followed by a discussion in section~\ref{discussion}, 
which presents more insight into our results. Section~\ref{related-work} 
describes related work and section~\ref{conclusion} presents conclusions and 
future work.

\section{Design Goals} ~\label{design}
We propose using the following design goals for sensor network protocols
for applications that have non-uniform information dissemination
requirements.

\begin{itemize}
\setlength{\itemsep}{-0.1em}
\item \emph{Energy efficiency}. As sensor nodes are
  battery-operated, protocols must be energy-efficient to maximize system
  lifetime.  

 \item \emph{Accuracy}.  Accuracy is a measure of the sensing
  fidelity: obtaining accurate information is the primary objective of
  a sensor network.  Accuracy is a metric that is application-specific
  both in terms of the appropriate metric and the required fidelity
  level. There is a trade-off between accuracy, latency and energy
  efficiency. In the applications we target, it is acceptable for sensors
  to have
  information with low accuracy about locations that are far away,
  but they should have highly accurate information about locations that
  are close by.
  Because of this non-uniform information
  dissemination requirement, a given sensor will not 
  have all the information from
  all other sensors at every point in time. Consider a case where
  sensor $S_{1}$ receives every $n^{th}$ packet from another sensor
  $S_{2}$. In this case, $S_{1}$ receives the $i^{th}$ packet from
  $S_{2}$ at time $t_{1}$ and the $(i+n)^{th}$ packet from $S_{2}$ at
  time $t_{2}$, ($n > 1$). Thus, in the interval ($t_{1}$,$t_{2}$),
  the information $S_{1}$ has about $S_{2}$ is not as accurate as a
  sensor that receives every update that $S_{2}$ sends.  We define
  accuracy in terms of the difference between the local value of the
  information and the actual value.

\item \emph{Scalability}. Scalability for sensor networks is also a
  critical factor. For large-scale networks, distributed protocols 
  are needed such that the protocol is based on localized interactions 
  and does not need global knowledge such as the current network topology.

\end{itemize}

With these design goals in mind, in this paper we present simple
deterministic protocols (Filtercast and RFiltercast) and
non-deterministic protocols (unbiased and biased protocols) to achieve
non-uniform information dissemination.  Compared to flooding, these
protocols reduce the cost of communication by reducing the number of
packet transmissions and receptions.  At the same time, these
protocols are designed to operate within the application-specific
tolerance in terms of accuracy.  Our results indicate that these
protocols outperform flooding in terms of energy efficiency by
trading-off accuracy for energy 
while keeping the accuracy acceptable by the application. The next
section describes the details of each of these protocols.

\section{Dissemination Protocols} \label{protocols} 

This section introduces the mechanisms of several protocols that
perform non-uniform information dissemination.  Similar to traditional
sensor networks, every sensor in the network serves as a source of
information to be spread throughout the network.  Unlike traditional
networks where a specific node serves as the \textit{sink} node, every
sensor in our system receives and stores some data from the other sensors in 
the system.  

We begin our protocol discussion with a traditional flooding
algorithm.  Flooding achieves \textit{uniform} information
dissemination, and serves as a baseline of comparison for the rest of
our protocols.  Following this, we introduce two new deterministic 
protocols and analyze two non-deterministic protocols~\cite{tilak-ms}. 

\subsection{Traditional Flooding}

In flooding, a sensor broadcasts its data, and this is received by all
of its neighbors. Each of these neighbor sensors rebroadcasts the
data, and eventually each sensor in the network receives the data.
Some memory of packets is retained at each sensor to ensure that the
same packet is not rebroadcast more than once.  If each sensor
broadcasts its data, then with this flooding protocol, every sensor in
the network will receive data from every other sensor.  Thus, ignoring
distribution latency, which is the amount of time required for a
packet to travel from the source to the farthest sensor in the
network,
every sensor has an identical view of the network at every point in
time (ignoring packet collisions and timing issues).
Furthermore, the protocol itself is simple and straightforward to
implement.  Unfortunately the simplicity and high accuracy come at the
price of high energy expenditure.  The massive data replication
requires active participation from every sensor in the network, and
thus sensors can quickly run out of energy.

\subsection{Deterministic Protocols}

In analyzing the flooding algorithm, it is apparent that to achieve
non-uniform information dissemination, one approach is to simply
have intermediate nodes forward fewer packets.  The two protocols
we introduce here, Filtercast and RFiltercast, achieve just that by
deterministic means.

\subsubsection{Filtercast}

As the name suggests, Filtercast filters information at each sensor
and does not transmit all the information received from other sensors
in the network.  Filtercast is based on a simple idea of sampling
information received from a given source at a certain rate $n$, specified as a
parameter to the protocol.  The lower the value of $n$, the more
accurate the information disseminated by the protocol.  When $n=1$,
Filtercast behaves identically to flooding.  During protocol
operation, each sensor keeps a count of the total number of packets it
has received so far from each source, $source_{cnt}$.  A sensor forwards
a packet that it receives from $source$ only if $(source_{cnt} \; mod
\; n) == 0$, then increments $source_{cnt}$.  We refer to the constant
$\frac{1}{n}$ as the filtering frequency.  The intuition behind this protocol
is that as the hop
count between a source node and a sink node increases, the amount of
information re-transmitted decreases due to the cascading effect of
the filtering frequency at each subsequent sensor.

While this reduces the total number of transmissions compared to
flooding, the state information maintained at each sensor increases.
Specifically, each sensor must maintain a list of all the sources it has
encountered from the start of the application and a count of the
number of packets seen from each of these sources.  As this increases
linearly with the size of the network, it may pose some scalability
problems. 

\subsubsection{RFiltercast}

One potential problem with Filtercast is the synchronization of the
packets transmitted by the neighbors. For example, consider a scenario 
where sensors $s_{2}$ and  $s_{3}$ are one-hop neighbors of both 
$s_{1}$ and  $s_{4}$, while  $s_{1}$ and  $s_{4}$ are two hops away from
each other. In this case, if Filtercast is used as a dissemination protocol,
then  $s_{2}$ and  $s_{3}$ will end up forwarding either all odd or all even 
packets (synchronized on forwarding the packets) from $s_{1}$ to $s_{4}$,
effectively transmitting redundant information.

Our intuition is that if we can
remove this redundancy, we may be able to increase the accuracy of the
protocol without increasing the energy expended.

To address this effect, we propose another protocol: Randomized
Filtercast (RFiltercast).  In this variant of Filtercast, the
filtering frequency $\frac{1}{n}$ is still the same for all sensors,
but each sensor generates a random number $r$ between $1 \ldots n$ and
re-transmits a packet if $(source_{cnt} \; mod \; n)-r == 0$.
Intuitively, this means that each sensor considers a window of size
$n$ and will transmit only one of the packets from a given source in
this window.  So, for a window of size $2$, half of the packets will
be selected for re-transmission, but instead of always re-transmitting
the first of the two packets (as in Filtercast), the sensors that choose 
$r=1$ will transmit the first of the two packets while the sensors that choose
$r=2$ will transmit the second of the two packets.

While our intuition was that the same energy would be expended by RFiltercast
as for Filtercast, this turns out not to be true.  In fact, RFiltercast
transmits more packets than Filtercast, but fewer than Flooding, putting its
energy expenditure in between the two. This effect happens because in
RFiltercast, a node receives more packets from a given source, as described
above, and thereby ends up transmitting more packets on behalf of the
source. Note that forwarding decisions at an intermediate node are not based 
on packet IDs but are based on the number of unique packets received from the 
source node. Therefore, due to the reception and transmission of more 
packets in 
RFiltercast, the energy dissipation of RFiltercast is higher than that of 
Filtercast.

While RFiltercast has more transmissions, increasing its energy expenditure, 
it also has improved accuracy over Filtercast.
The crucial point to extract is that RFiltercast should, on average, propagate
information faster than Filtercast, leading to more accurate data 
throughout the network, but RFiltercast will require less energy than flooding.

\subsection{Randomized Protocols}
Both RFiltercast and Filtercast are lightweight and easy to analyze due to 
their deterministic nature.  However, they still have some overhead in terms
of the state required at each node.

We next describe several probabilistic protocols.
In these protocols, when a sensor receives a packet, it chooses a
random number and then decides whether to forward the packet or not
based on the number chosen.
We  classify these protocols into two categories:
biased and unbiased protocols.  In the biased protocol, sensors bias
their decision about whether to forward a packet based on the location
of the source, where packets from close sensors are more likely to be
forwarded than packets from distant sensors.  In the unbiased
protocol, all packets are forwarded with equal probability.

\subsubsection{Unbiased Protocol}

The notion of using probabilities to flood packets throughout a
network has been studied previously
~\cite{barrette-03,FLDDRFT,dream,probflood}, but to the best of 
our knowledge, no studies exist that explore its applicability to non-uniform
information granularity requirements.  Similar to the deterministic
protocols, the unbiased protocol also takes a parameter that affects
the accuracy of the forwarding.  In this case, the parameter specifies
the probability that a packet should be forwarded.  In the case of
the unbiased protocol, this value is the same for each incoming packet.

The main advantage of this protocol is its simplicity and low
overhead. As every packet is forwarded only with a certain
probability, the protocol results in less communication compared to
flooding (proportional to the forwarding probability).  Also, the
protocol does not require state to be kept, giving this protocol
the potential to scale well.

To adjust the accuracy of the information throughout the network, the
forwarding probability can be tuned according to the application
needs.  The primary tradeoff, however, is energy for accuracy.  In
general, as the forwarding probability increases, the behavior
converges toward flooding.  While our current study considers only
constant probabilities, in the future we will look at the possibility
of probabilities being adjusted dynamically to adapt to the current
network traffic and the application needs.

\subsubsection{Biased Protocol}

In this protocol, the forwarding probability is chosen to be
inversely proportional to the distance the packet has traveled since
leaving the source sensor.  In other words, if a sensor receives a
packet from a close neighbor, it is more likely to forward this than a
packet received from a neighbor much farther away.  To estimate
distance between sensors, a sensor examines the TTL (time-to-live)
field contained in the packet.  If we assume all sensors use the same
initial TTL, we can use the current TTL to adjust the forwarding
probability for each packet. The following tuples indicate the forwarding 
probabilities used (second number in the tuple) when the packet has traveled
the number of hops in the range specified in the first part of the tuple: 
$<1-3,~0.8>$, $<4-6,~0.6>$, $<7-9,~0.4>$, $<10+,~0.2>$.\label{biasedprob}

Using TTL to estimate distance is simple; however, note that TTL does
not always indicate the exact distance between two sensors.  For
example, consider a source node $S$ and a destination node $D$.  It is
possible that either due to congestion or collisions, a packet gets
dropped along the shortest path and another packet reaches node $D$
via a longer route. In that case, the TTL would give a false estimate
of distance. However, in a static network, node $D$ can always
maintain its current estimate of the TTL to node $S$.
In the case of mobile sensors, as the distance between $S$ and $D$
changes (decreasing, for example, as the nodes get closer), this is reflected in 
the subsequent packets (higher TTL value) and thus node $D$ gets more and
more accurate information about $S$. We use
the TTL-based approach for the biased protocol mainly for its
simplicity, resilience to mobility and energy efficiency.

Similar to the unbiased protocol, this biased protocol requires no additional
storage overhead unless node distances are stored, and the protocol itself is 
completely stateless
(note, however, that this does not eliminate the caching of recently
seen packets in order to avoid re-broadcasting the same packet
multiple times).  Therefore, this protocol scales as well.

\section{Experimental Study}  \label{experimentation}

In order to analyze the protocols, we use the ns-2 discrete event
simulator~\cite{ns-2}.  Table~\ref{tab:params} lists the relevant
parameters used during our simulations.

\begin{table}[h]
\begin{center} 
\caption{Simulation parameters.}
\label{tab:params}
\vspace{0.1in}
\begin{tabular} {|l|c|} \hline
Simulation area         & $800\times800~m^2$ \\ \hline
Transmission range      & $100~m$ \\ \hline
Initial Energy          & $10000~J$  \\ \hline
MAC Protocol            & $802.11$ \\ \hline
Bandwidth               & $1~Mbps$      \\ \hline
Transmit Power          & $0.660~W$ \\ \hline
Receive Power           & $0.395~W$ \\ \hline
Idle Power              & $0.0~W$ \\ \hline
Number of Nodes         & $100$ \\\hline
\end{tabular}
\end{center}
\end{table}

In the case of static networks, we consider two sensor deployment
strategies: uniform and random.  In a uniform deployment strategy,
sensors are distributed with some regular geometric topology (e.g., a
grid).  With random deployment, sensors are scattered throughout the
field with uniform probability. For a battlefield-like scenario, random
deployment might be the only option, but with applications such as
animal tracking in a forest, sensors may be deployed in a deliberate,
uniform fashion.

In order to simulate sensor readings, we divide the simulation into an
initialization phase and a reporting phase. During the initialization
phase, each sensor chooses a random number between $0$ and $100$ to
serve as its initial sensor reading.  During the reporting phase, each
sensor increments its reading by a fixed amount at fixed intervals.
In the real world, due to correlation among physically co-located
sensors, sensors will have a different reading pattern;
however, this simulation does provide us with valuable information
about the behavior of our protocols under various conditions. In the
latter part of this section, we present a revised data model that
tries to capture correlation among sensor readings. Our results
indicate that the overall behavior of the protocols shows a very similar
trend for both data models.

\subsection{Traffic Load Study}

This study focuses on evaluating the effect of a change in traffic
load for both grid and random topologies. In the first set of
experiments, we study the effect of varying traffic loads
systematically from $5$ packets/sec to $1$ packet/ $2$ sec. The goal
of these experiments is to understand the relationship between
accuracy, reporting rate, and network capacity for both uniform and
non-uniform dissemination scenarios.

Note that in order to calculate accuracy, we find the difference
between a sensor's local view of another sensor's data and the actual value
of that sensor's data.  A \textit{view} is essentially the latest data 
that one sensor has from another sensor. This view is then
normalized based on distance. Let $R(S_{i,j})$ denote sensor $S_i$'s view 
of sensor ${S_j}$'s data, and let $n$ be the total number of sensors in the 
network.
%and $\gamma$ was set to 100 (transmission range).
The weighted error $e_{i}$ for a sensor $S_{i}$ is given as:

\begin{equation}
 \label{erroreqn1}
 e_{i} = \frac{1}{n} \, \Sigma_{j=1, j \neq i}^{n} | (R(S_{i,j}) - R(S_{j,j}) | * \,w_{ij} 
\end{equation}

\begin{equation}
 \label{erroreqn2}
% w_{ik} = 1 / eucdist(S_{i},S_{k})
 w_{ij} = \frac{1}{\frac{\lceil d(S_{i},S_{j}) \rceil}{\gamma}}
\end{equation}
where $d(S_i,S_j)$ is the Euclidean distance between sensors $S_i$ and
$S_j$ and $\gamma$ was set to 100 (the transmission range of each node).
The fist equation shows that for a given sensor we calculate weighted
average error with respect to all other sensors in the network. We vary the 
weight in terms of distance with a step size of 100 meters. 

Note that Euclidean distance is used as the weighing
factor so that the higher the distance, the smaller the contribution
of error toward overall error. This error calculation describes our
non-uniform data dissemination requirement by giving higher weight to
errors for data that originated in a close neighborhood and lower
weight to errors for data that originated from a distant sensor.  It
is worth noting that although we refer to this as \textit{error},
because the value of the data at the source increases linearly, it
also represents the accuracy of the data.

Our results indicate that with flooding, congestion is a severe
problem, and other protocols are less prone to the congestion problem.
In this type of application, the effect of congestion is worse than
that observed in traditional sensor networks~\cite{tilak-wsna}.  From
the simulation studies, we can see that flooding is the
least energy-efficient protocol and has the highest error if the
network is congested. RFiltercast and the biased protocol are more
energy-efficient than flooding and provide low error in most cases.
Filtercast and the unbiased protocol are the most energy-efficient
protocols, but their accuracy is good (low error) only at higher
sending frequencies.

\subsubsection{Grid Topology}

\begin{figure*}
\begin{center}
\mbox{
\subfigure[Data rate 5 packets/sec.\label{staticgrid5}]{\epsfig{file=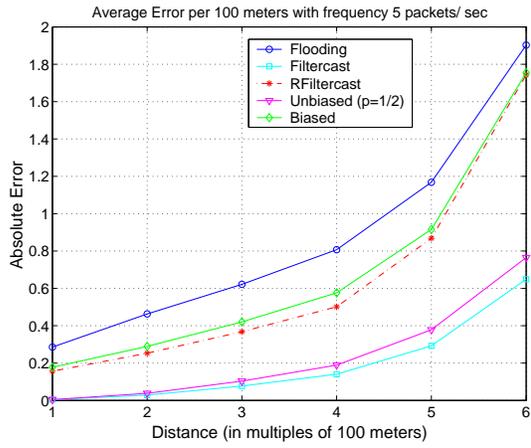,scale=0.4,silent=}} \quad
\subfigure[Data rate 2 packets/sec.\label{staticgrid2}]{\epsfig{file=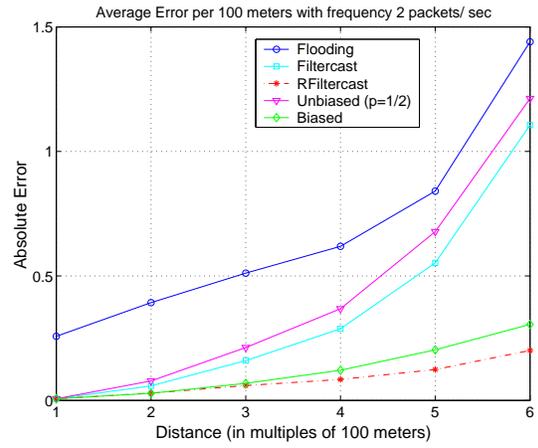,scale=0.4}}
} 
\mbox{
\subfigure[Data rate 1 packet/sec.\label{staticgrid1}]{\epsfig{file=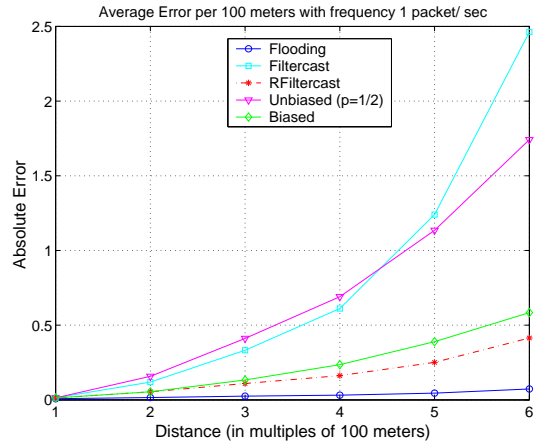,scale=0.4}}\quad
\subfigure[Data rate 1 packet/ 2 sec.\label{staticgridlow}]{\epsfig{file=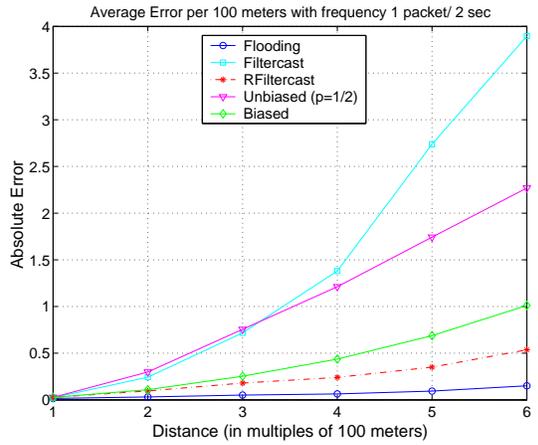,scale=0.4}}
}
\caption{Grid Topology: Mean absolute error as a function of distance for different source data rates.}
\label{staticgrid} 
\end{center}
\end{figure*}

Figure~\ref{staticgrid} shows the performance of flooding, Filtercast,
RFiltercast, and the biased and unbiased randomized protocols under
various traffic loads for the grid topology. In these graphs, distance
is varied across the X-axis (in steps of 100 meters) and the Y-axis shows mean 
unweighted error (mean absolute error). Note that, with non-uniform information dissemination,
as the distance between the source node and sink node increases, loss in information
precision is acceptable.

From Figure~\ref{staticgrid5}, where the data rate is $5$ packets/sec,
we can see that even though theoretically flooding should have no
error, due to congestion, flooding has the highest error. This is due
to the fact that if the total traffic exceeds the network capacity,
congestion causes packets to be dropped and this gives rise to loss of
information and high error. At the same time, high traffic results in
higher collisions. In this situation, even RFiltercast and the biased
randomized protocol result in high traffic load and thus they have
high error as well. However, both Filtercast and the unbiased
randomized protocol (with forwarding probability of 0.5) perform well
in this case because the traffic load does not exceed the available
network capacity.  As expected, for all protocols the error increases
as the distance from the source increases, resulting in non-uniform
information across the network.

When the sending frequency is changed to $2$ packets/sec, as shown in
Figure~\ref{staticgrid2}, flooding the network still causes congestion
and thus flooding has high error. However, now for both RFiltercast
and the biased protocol, the load does not exceed the network capacity
and their performance is better than in the previous case. Also, note
that now these two protocols perform better in terms of error rate
than the unbiased protocol and Filtercast because of the fact that
they disseminate more information yet the information disseminated
does not exceed the network capacity.

When the sending frequency is lowered to $1$ packet/sec, as shown in
Figure~\ref{staticgrid1}, then even flooding does not exceed network
capacity. Since the network is no longer a bottleneck, flooding
disseminates the maximum information successfully and clearly has the
lowest error.  Both the biased and RFiltercast protocols perform
better than the unbiased protocol and Filtercast.  The unbiased
protocol and Filtercast have the highest error in this case because
they do not disseminate as much information as the other protocols.
The same trend continues even for the lowest sending frequency, shown
in Figure~\ref{staticgridlow}.

The interesting point about these results is the oscillatory
behavior of the energy-error curves. To elaborate further on this, if
the total data exceeds network capacity, then any further data on the
channel will increase congestion and decrease overall lifetime of the
network.  When the amount of data transmitted is below network
capacity, then there is a trade-off between energy spent and accuracy
observed. This is because as long as the total data does not exceed
network capacity, sending more data will improve accuracy at the cost
of energy spent in communication.  However, with non-uniform
information granularity, accuracy between two sensors is proportional
to distance between them. Therefore, RFiltercast and Filtercast try to
achieve this by filtering packets and the randomized protocols try to
achieve this by probabilistically forwarding packets.

\begin{figure}[t] 
\centerline{\epsfig{file=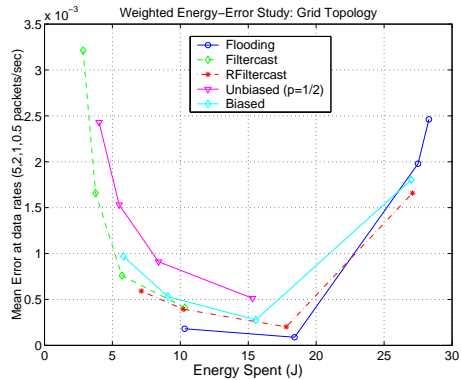,width=0.75\linewidth,silent=}}
\caption{Grid: Weighted energy-accuracy tradeoff.}\label{werrenggridstatic}
\end{figure}

Figure~\ref{werrenggridstatic} shows the trade-off between energy and
weighted error,
using the weighted error calculation method described in
Eqns.~\ref{erroreqn1} and \ref{erroreqn2}.
In Figure~\ref{werrenggridstatic}, the 
X-axis indicates the energy spent in Joules and the Y-axis shows mean weighted 
error. Each point represents one of the sending frequencies, ranging from 5 packets per second for the left-most point of each curve to one packet per two seconds for the right-most.

For flooding, when the reporting frequency is highest, the energy
spent is maximum.  However, as mentioned earlier, congestion and
collisions cause high error. As the sending frequency decreases to the
point that total traffic does not exceed network capacity, the error
also decreases. Flooding performs the best in terms of accuracy
(minimum error) when the sending frequency is $1$ packet/sec; after
this rate, the error starts increasing due to the fact that not enough
information is propagated. This is an interesting phenomenon, where
the error oscillates between these two bounds. The upper bound is a
function of the network capacity, whereas the lower bound is a
function of the application-specific accuracy.  Previous research has
also shown this phenomenon~\cite{tilak-wsna}.

Based on the energy-error trade-off, we can say that at high sending
frequency, flooding performs the worst by spending high energy while
not providing accurate information (high error).  RFiltercast and the
biased protocol start performing better than flooding at high rates.
There is a considerable difference between energy and error for
RFiltercast and the biased protocol compared to flooding at the
sending frequency of $2$ packets/sec.  As one can anticipate, flooding
performs better than all other protocols in terms of accuracy when the 
sending frequency is $1$ packet/sec, but note that there is not much 
difference between
flooding, RFiltercast and the biased protocol even when the network is
operating in the non-congested mode. Filtercast and the unbiased
protocol perform best in terms of energy and error at high sending
frequencies and their performance relative to the other protocols
starts to degrade as the sending frequency is reduced.

The desirable mode of operation for a protocol is in the region where
minimum energy is spent and low error is observed. Note that the
desired mode of operation for a protocol depends on factors such as
network density, transmission range of the radios, etc.  In our future
work, we will perform an analytical study to address this issue.  From
Figure~\ref{werrenggridstatic}, this zone lies around sending
frequency $2$ packets/sec to $1$ packet/sec for RFiltercast and the
biased protocol, whereas for flooding it lies at sending frequency $1$
packet/sec.
We want to point out that as the network size increases, flooding
can pose severe problems in terms of scalability and energy
efficiency.  Therefore, randomized protocols should be considered as
viable alternatives in these cases. In our experiments we had a
network of 100 sensors, but with a network of thousands of sensors we
believe that randomized protocols will perform much better than
flooding.
With randomized protocols, the biased protocol performs the best by
spending moderate energy and getting high accuracy.

Our results show that randomized protocols can achieve high energy
savings while at the same time achieving acceptable accuracy with
almost no overhead.  Also note that RFiltercast and the biased
protocol have almost equivalent error curves while the biased protocol
has negligible overhead.

\subsubsection{Random Topology}
\begin{figure*}
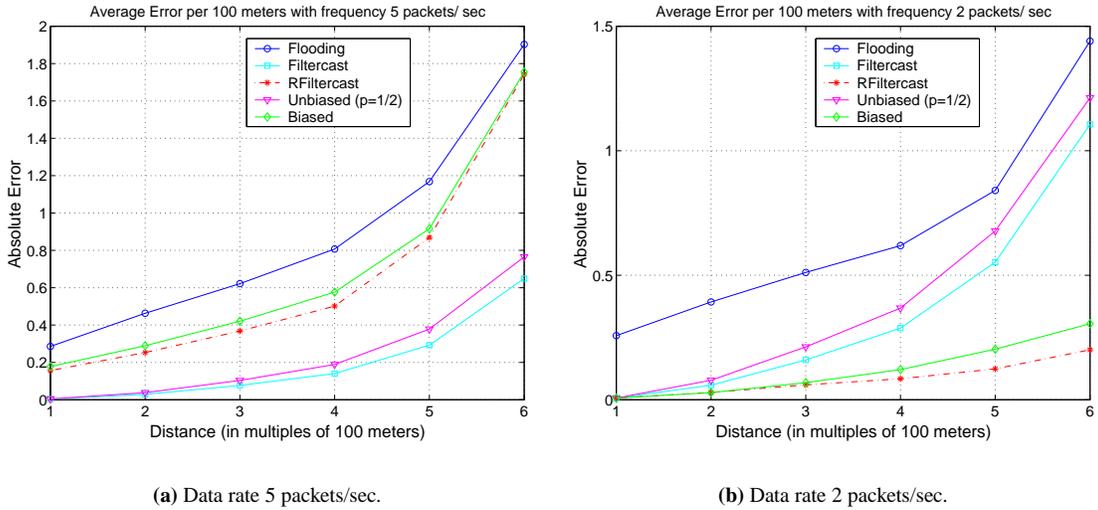

\begin{center}
\mbox{
\subfigure[Data rate 5 packets/sec.\label{staticrand5}]{\epsfig{file=figs/staticgridf5.eps,scale=0.4}} \quad
\subfigure[Data rate 2 packets/sec.\label{staticrand2}]{\epsfig{file=figs/staticgridf2.eps,scale=0.4}}
}

\caption{Random Topology: Mean absolute error as a function of distance for different source data rates.}
\label{staticrandom} 
\end{center}
\end{figure*}                                                                   Figure~\ref{staticrandom} shows our results with a random topology and the 
same traffic loads as before. The results for 1 and 2 packets/sec and the 
energy tradeoff study show expected results similar to those achieved for 
the grid topology, and are therefore not shown here.
It is not clear whether regular deployment will offer
advantages over uniformly distributed random deployment; if it does not,
random deployment is preferable because of its low cost.

\subsubsection{Transmission Range}

\begin{figure*}
\begin{center}
\mbox{
\subfigure[Data rate 5 packets/sec.\label{trstaticgridf5}]{\epsfig{file=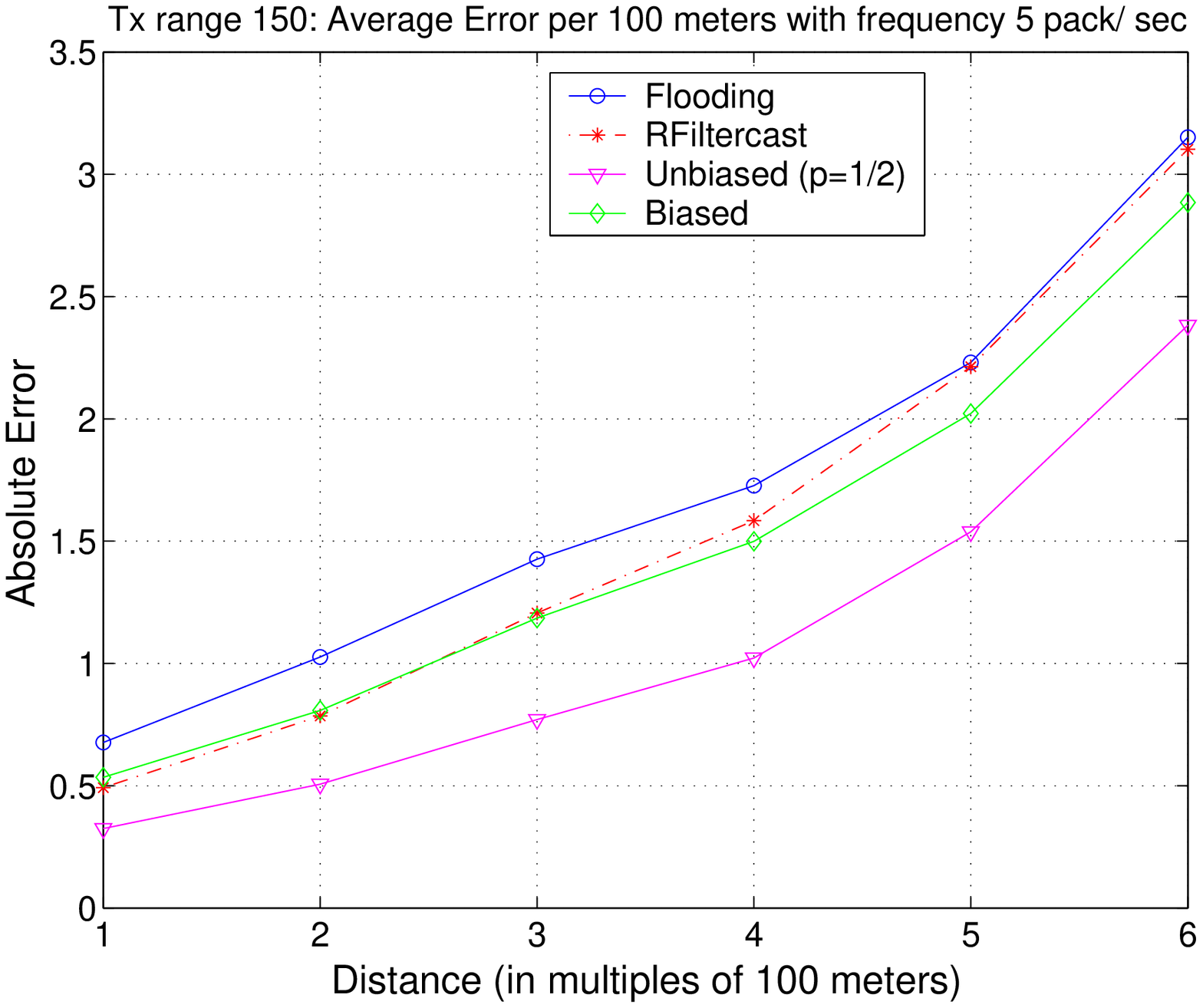,scale=0.4}} \quad
\subfigure[Data rate 2 packets/sec.\label{trstaticgridf2}]{\epsfig{file=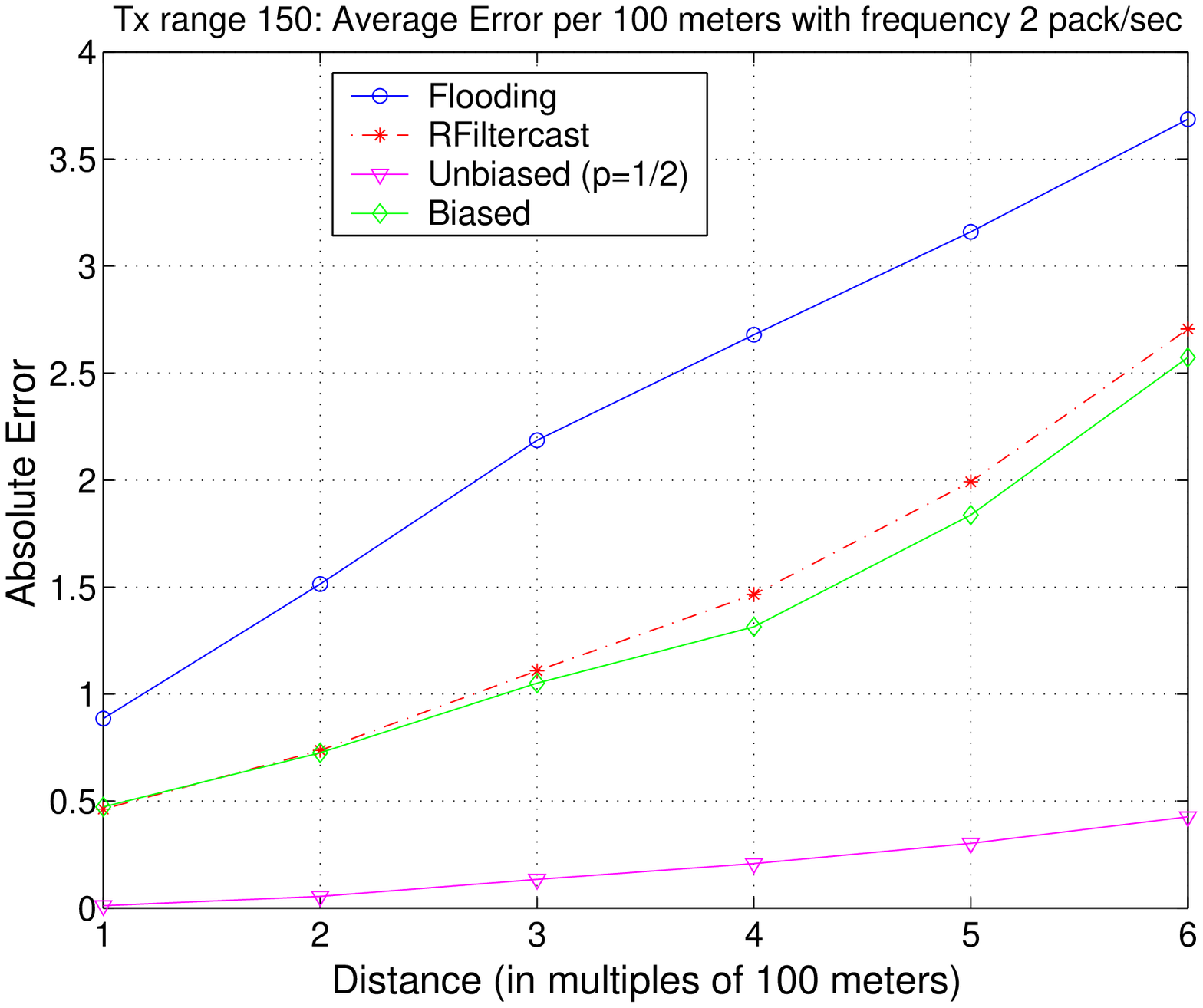,scale=0.4}}
}
\mbox{
\subfigure[Data rate 1 packet/sec.\label{trstaticgridf1}]{\epsfig{file=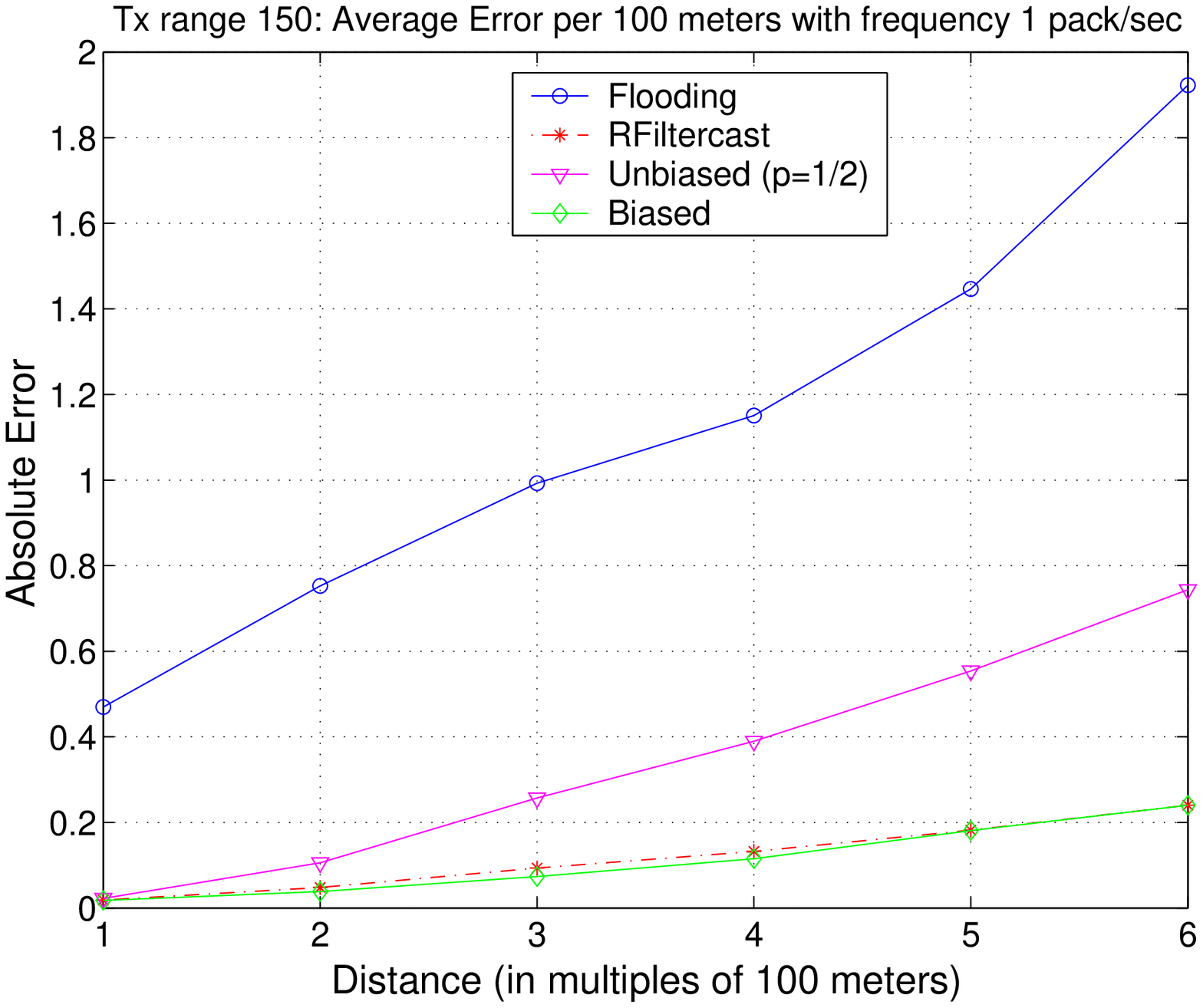,scale=0.4}}\quad
\subfigure[Data rate 1 packet/ 2 sec.\label{trstaticgridflow}]{\epsfig{file=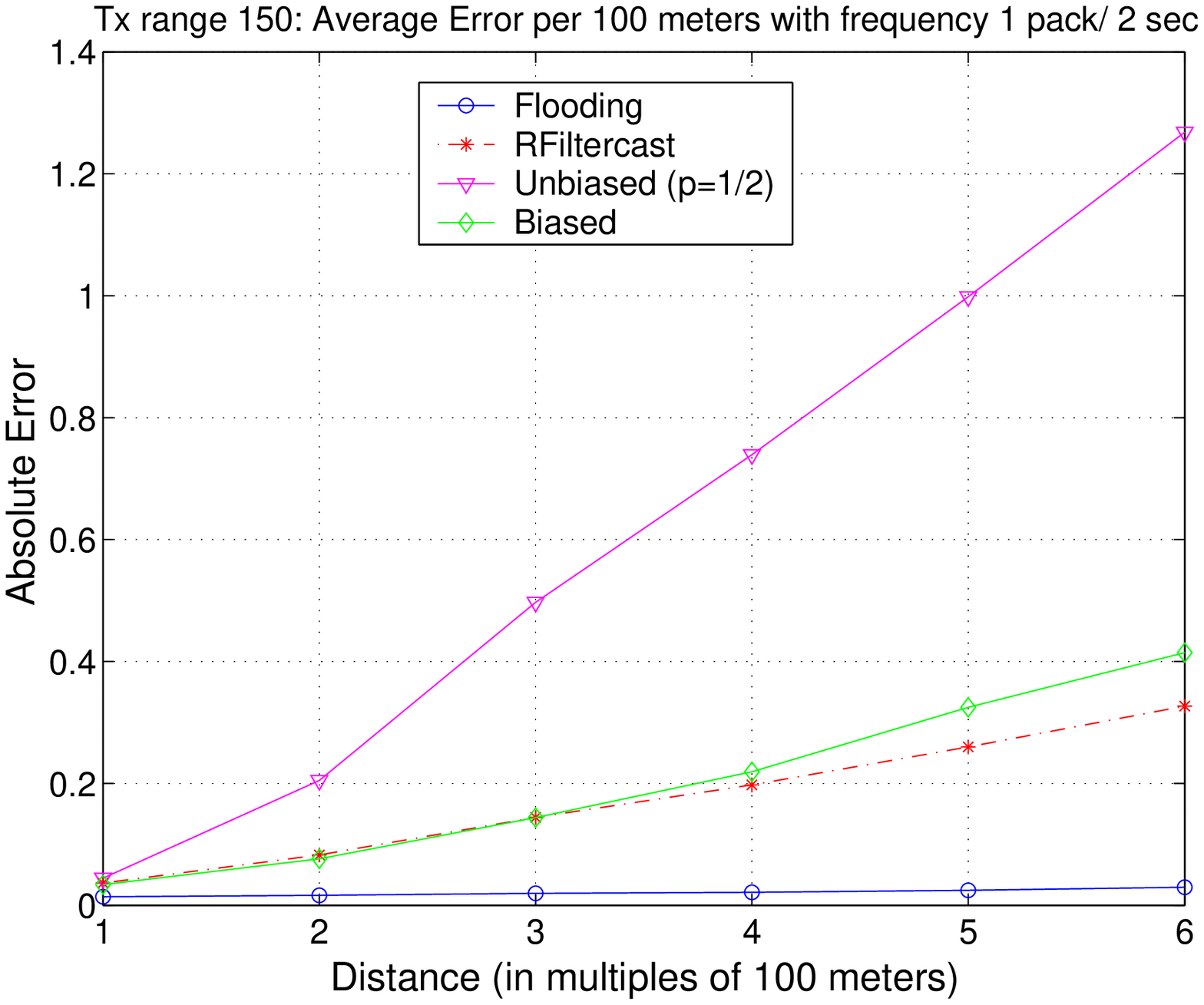,scale=0.4}}
}
\caption{Tx= 150 m (Grid Topology): Mean absolute error as a function of distance for different source data rates.}
\label{trstaticgrid} 
\end{center}
\end{figure*}                                                                   
Our next set of experiments show the effect of an increase in the transmission
range from the original 100 meters to 150 meters, while keeping the original 
10x10 grid topology. An increase in transmission range corresponds to
an increase in the degree (connectivity) of a sensor.  
This results in decreasing the capacity of the network, meaning
congestion occurs even at low sending frequencies. Intuitively, this
will make the overall situation worse if the network is operating in a
congested mode. This can be seen from our results, comparing
Figures~\ref{staticgrid5} and \ref{trstaticgridf5}, as there is an
increase in overall error for flooding, RFiltercast and the randomized
protocols.

In this case, even at the low sending frequency of $1$ packet/sec
shown in Figure~\ref{trstaticgridf1}, flooding does not perform well
due to network congestion. Previously, when the transmission range was
$100$ meters, flooding performed well at this sending frequency (see
Figure~\ref{staticgrid1}). However, when the network is not congested,
then due to the higher connectivity and shorter average hop length,
the average error decreases. For example, for RFiltercast and the
biased protocol, when the sending frequency is $1$ packet/sec, then
the maximum absolute error values with a transmission range of $100$
meters are $0.4$ and $0.6$ respectively, as shown in
Figure~\ref{staticgrid1}.  With the transmission range changed to
$150$ meters, Figure~\ref{trstaticgridf1} shows that the maximum
absolute error for RFiltercast and the biased protocol changes to
$0.24$ for both.

Similarly, all the protocols have low error values at a sending
frequency of $1$ packet/ $2$ sec when the transmission range is $150$
meters, as shown in Figure~\ref{trstaticgridflow}, compared to the
simulations with $100$ meters transmission range, shown in
Figure~\ref{staticgridlow}. 

Up to this point our study considered
static networks.  In the next subsection we analyze protocols for
non-uniform information dissemination in the
presence of mobility along with a revised data model.

\subsection {Mobility Study}

 To motivate the case for mobile sensors, consider a battlefield
scenario, where soldiers and armed vehicles are moving carrying tiny
sensors along with them. Each sensor is collecting information about
air contaminants so as to find out about potential biological/chemical
attacks. In this case as a soldier moves around, the
presence of an air contaminant sensed by the sensor changes depending
upon the sensor's current location in the battlefield. Also, 
for a small change in location, there is not a very high change in the
percentage of contaminant reported. We can think of this as a
spatio-temporal process, where there is both spatial and temporal
correlation among the readings reported by the sensors.  
This means that correlation among the sensor readings is a function of
the distance between them; the closer the sensors are, the higher the
correlation between their data. 

In order to model this application, we divide the simulation area of
$800$x$800$ meters into 16 squares, each called a zone. We assume that the
sensor readings (contaminant in this case) follow a normal distribution
in space (inter-zonal distribution).  For the inter-zonal distribution, we
set the mean to 20 and the standard deviation to 2. This corresponds to a
loose correlation among data sensed by all the sensors in the
battlefield.  Also, there will be very high correlation among data
reported by sensors within the same zone. We modeled the correlation among
sensors within a zone (intra-zonal) to follow a normal distribution but
with low variance compared to that of the inter-zonal distribution.  For
the intra-zonal distribution we set the standard deviation to be a random
number in the interval of $0$ to $0.5$ (both inclusive).  Also, we
vary the mean of the intra-zonal distribution as the simulation
progresses to reflect the temporal variations. During the initial half
of the simulation, the mean slowly increases and then during the later half
of the simulation it decreases gradually. For evaluating the
performance of these protocols, we calculate the weighted error  
in the same way that we did for the static networks (e.g., 
Eqns.~\ref{erroreqn1} and \ref{erroreqn2}). 
We want to emphasize that these zones are just an artifact of modeling 
the phenomenon (contaminant 
presence across the field). 

\begin{figure*}
\begin{center}
\mbox{
\subfigure[Data rate 5 packets/sec.\label{sp2f5}]{\epsfig{file=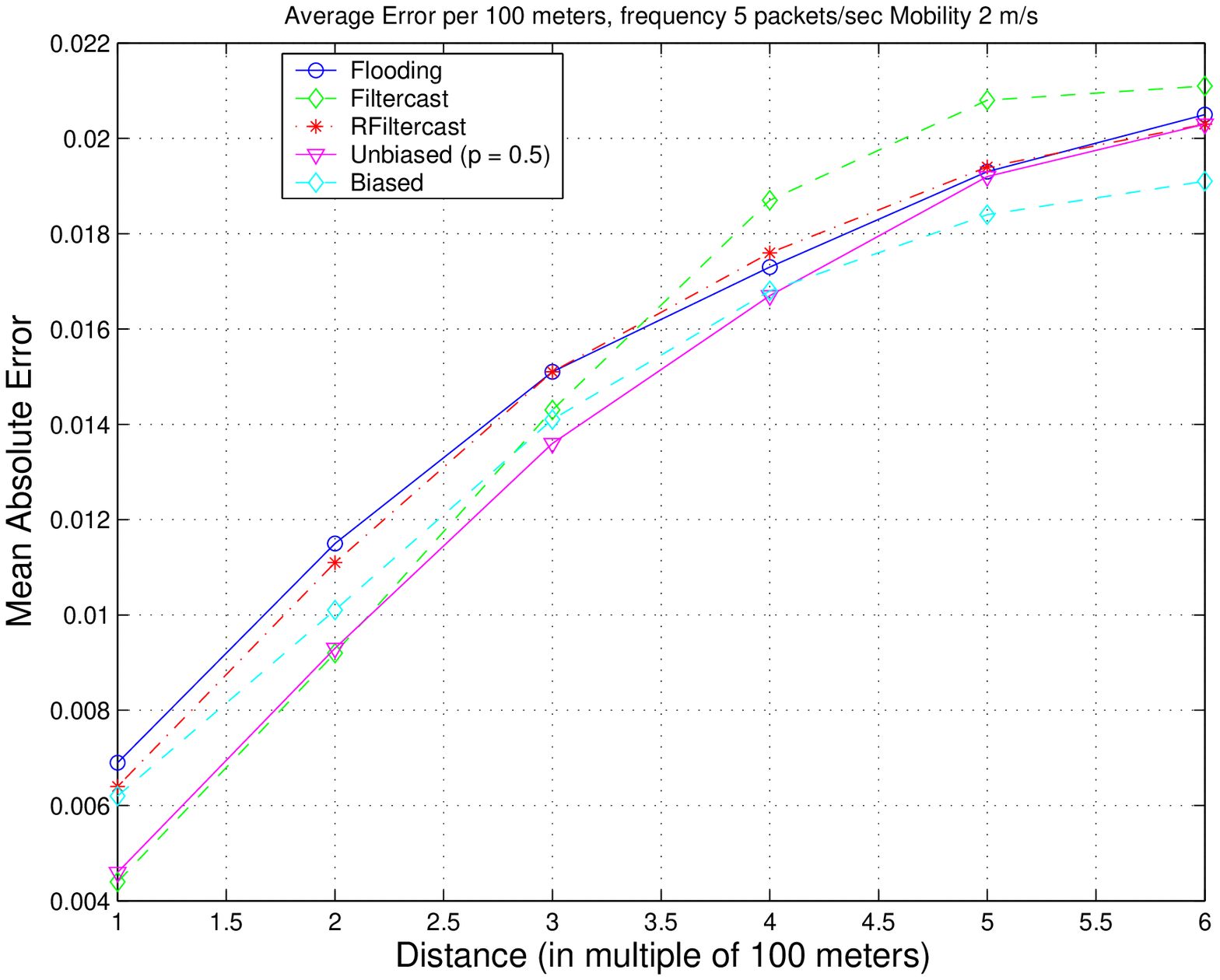,scale=0.4}} \quad
\subfigure[Data rate 2 packets/sec.\label{sp2f2}]{\epsfig{file=figs/speed-2-f5-error.eps,scale=0.4}}
}
\mbox{
\subfigure[Data rate 1 packet/sec.\label{sp2f1}]{\epsfig{file=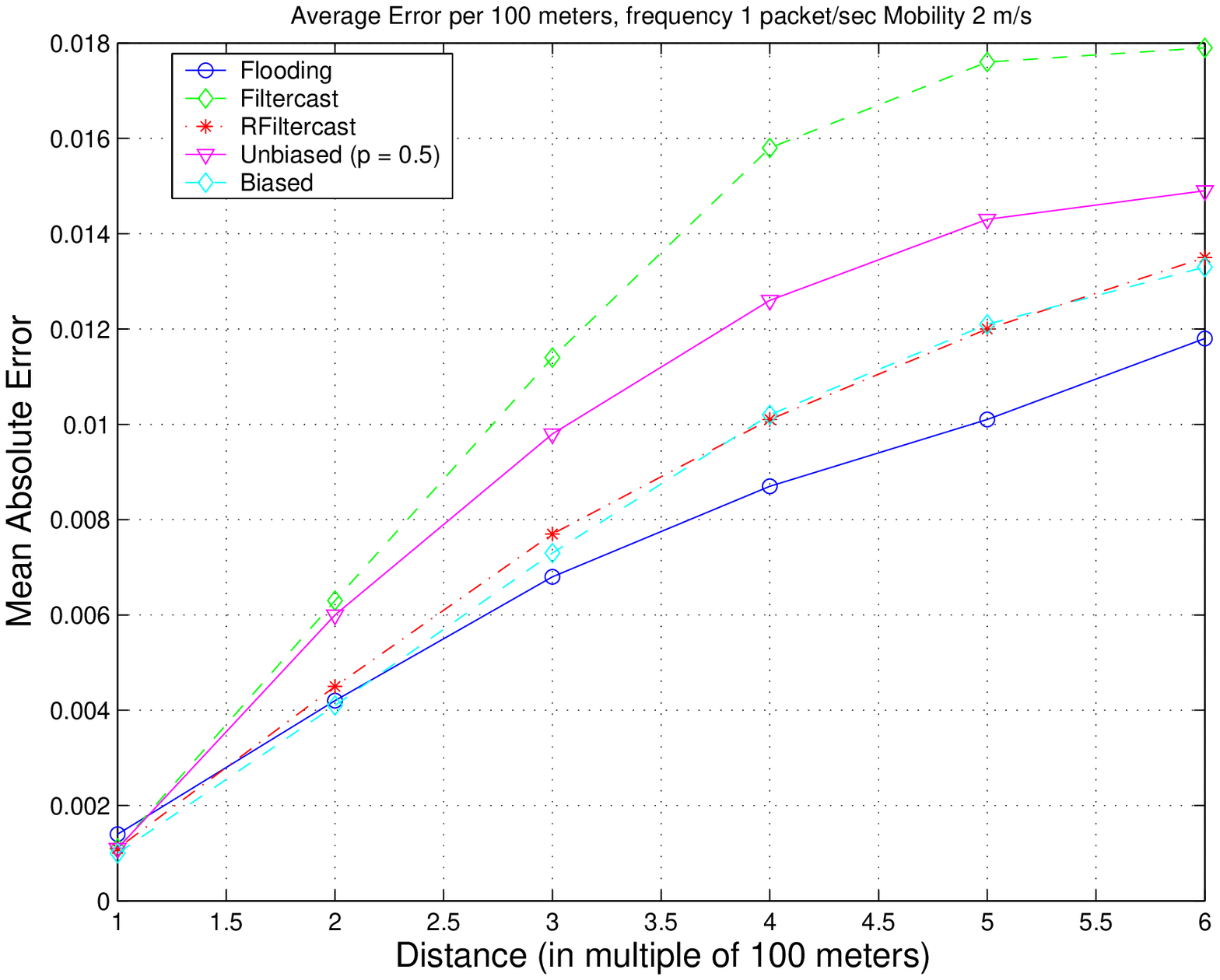,scale=0.4}}\quad
\subfigure[Data rate 1 packet/ 2 sec.\label{sp2flow}]{\epsfig{file=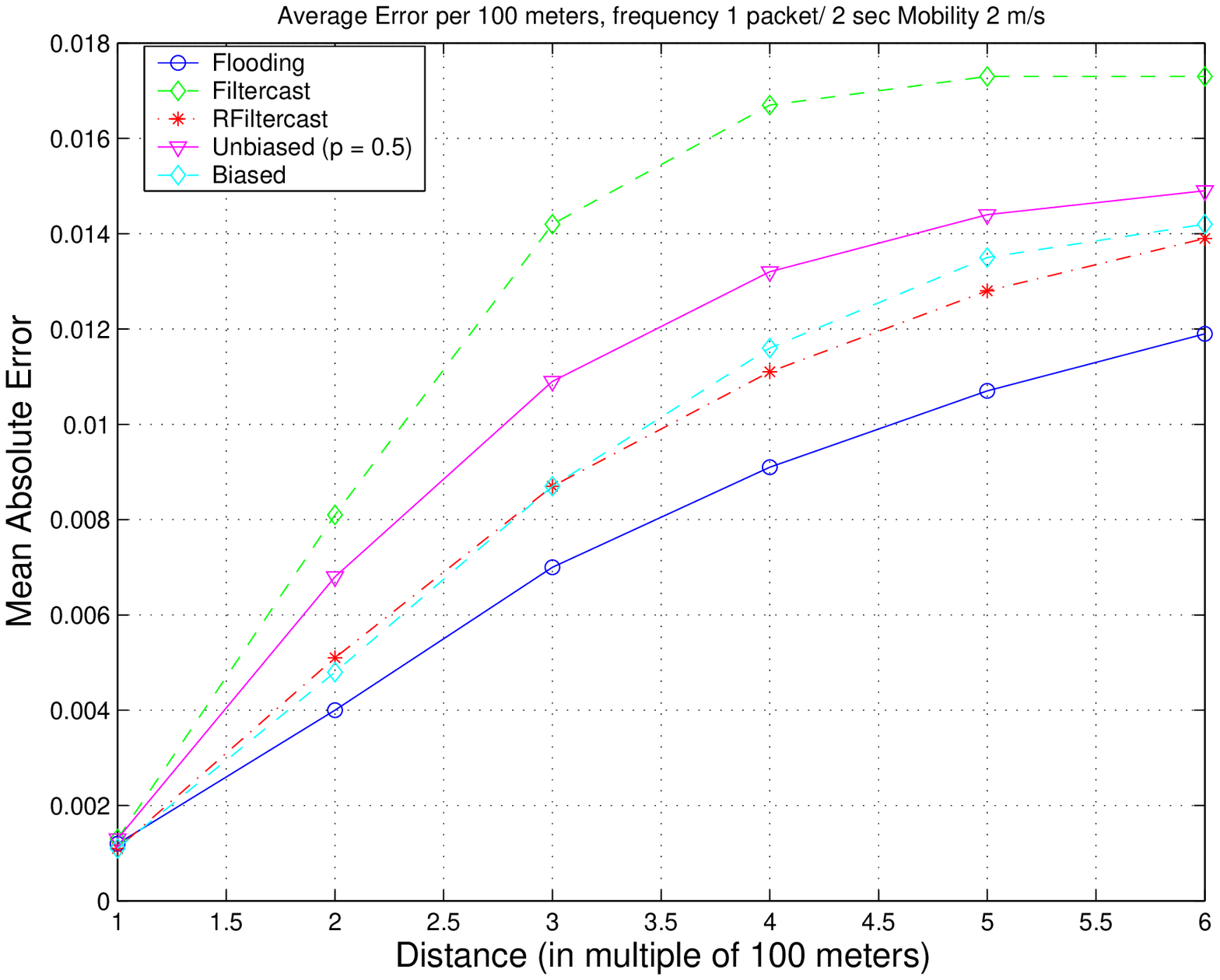,scale=0.4}}
}
\caption{Mobile sensors (speed 2 m/sec): Mean absolute error as a function of distance for different source data rates.}
\label{mhuman} 
\end{center}
\end{figure*}              

\begin{figure*}
\begin{center}
\mbox{
\subfigure[Data rate 5 packets/sec.\label{sp10f5}]{\epsfig{file=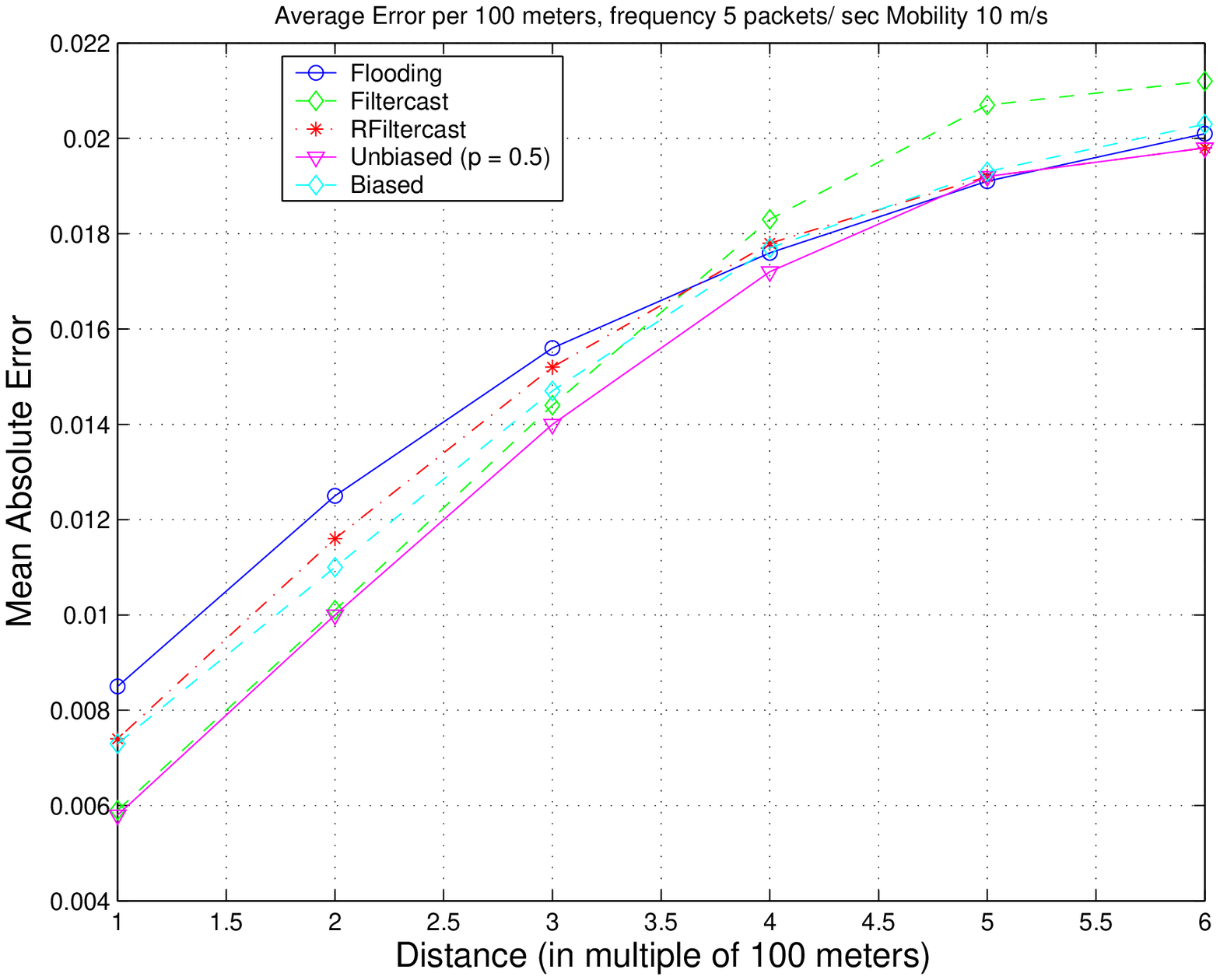,scale=0.4}} \quad
\subfigure[Data rate 2 packets/sec.\label{sp10f2}]{\epsfig{file=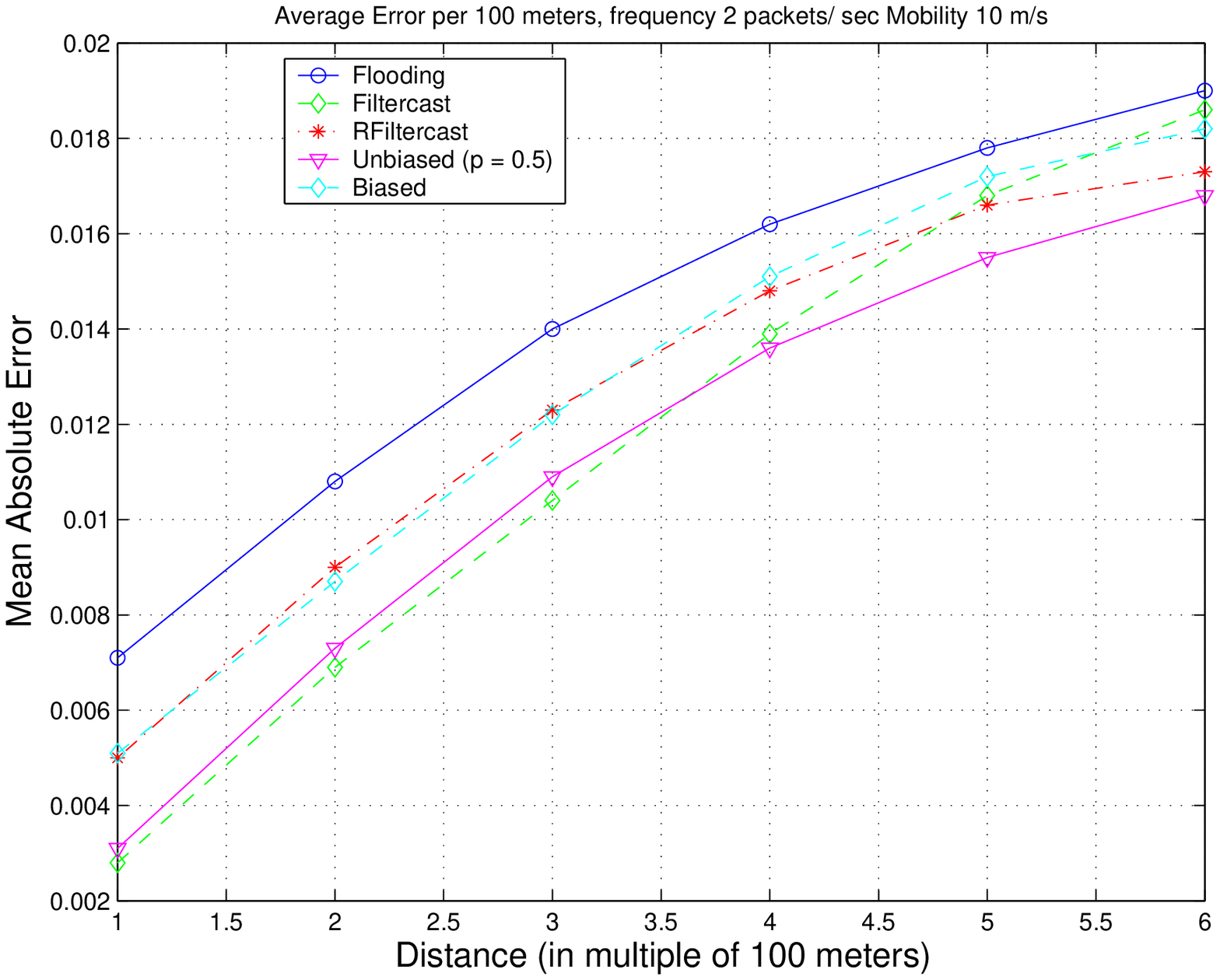,scale=0.4}}
}
\mbox{
\subfigure[Data rate 1 packet/sec.\label{sp10f1}]{\epsfig{file=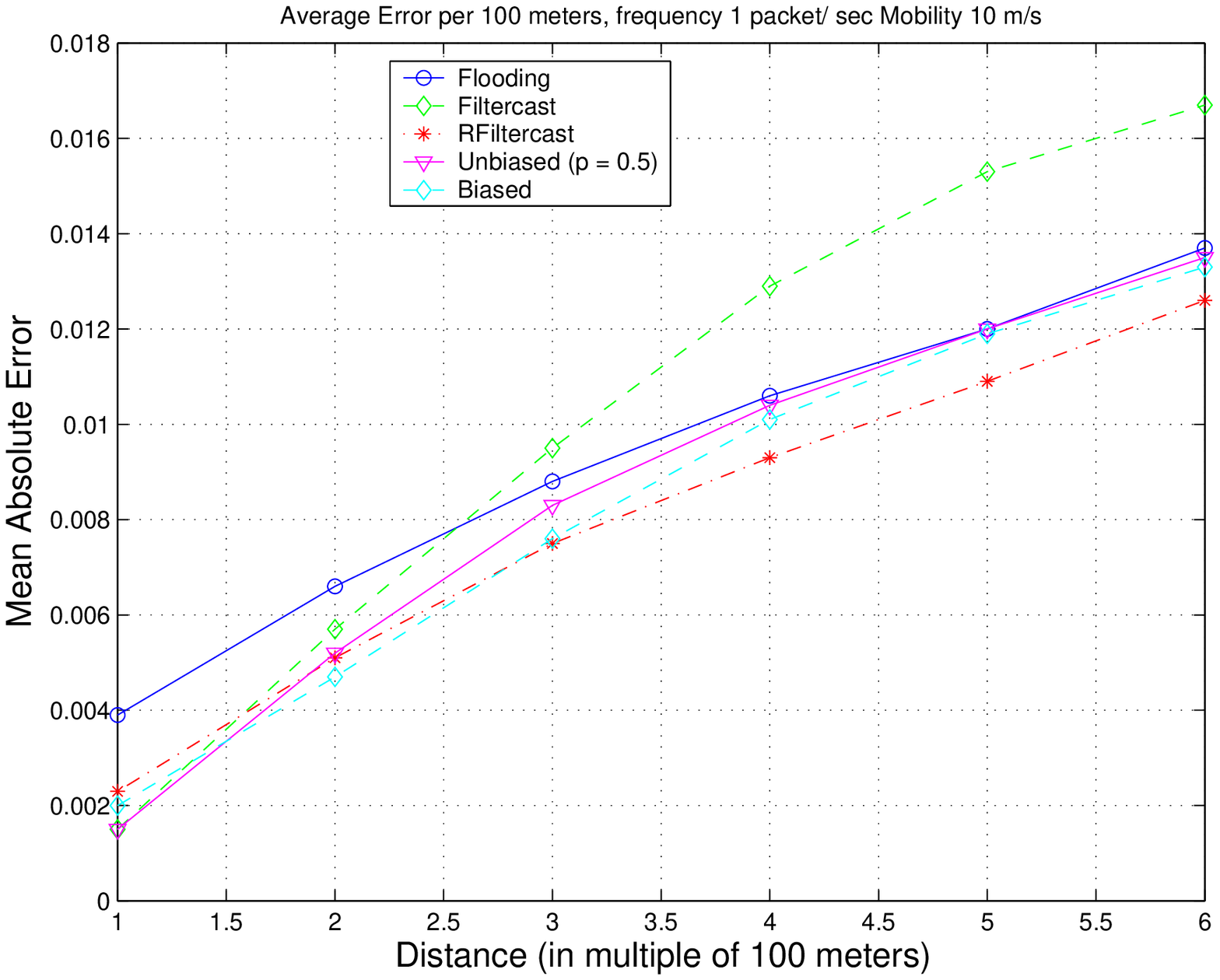,scale=0.4}}\quad
\subfigure[Data rate 1 packet/ 2 sec.\label{sp10flow}]{\epsfig{file=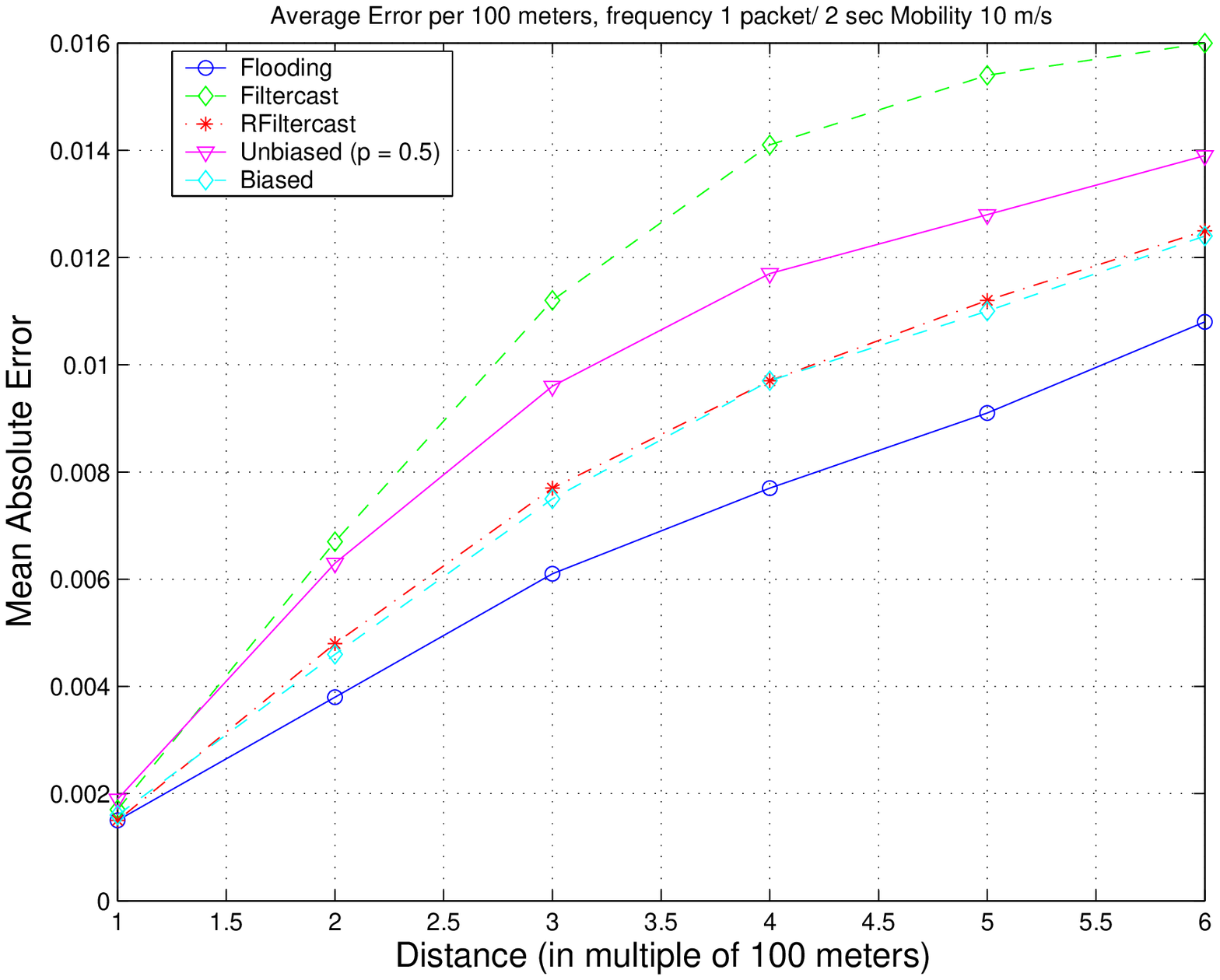,scale=0.4}}
}
\caption{Mobile sensors (speed 10 m/sec): Mean absolute error as a function of distance for different source data rates.}
\label{mvehicle} 
\end{center}
\end{figure*}     

For static networks, distance between every pair of sensors is fixed
(time invariant). However, in the case of mobile sensors, as the
sensors move around, the distance between a pair of sensors
changes. By non-uniform information granularity, we intuitively mean
that a sensor should have very precise information about its local
neighborhood and loss in accuracy should be proportional to the
distance between source and sink sensors. However, with mobile
sensors, as the sensors move, the local neighborhood of a sensor
changes as a function of time.  It is thus interesting to
study how the protocols (both deterministic and randomized) react
to these neighborhood changes.

We now analyze the performance of all the protocols when nodes are mobile with
the revised data model.  For mobility we consider the following two
cases. In the first case we set the maximum speed of the sensors to 2
m/s (Figure \ref{mhuman}) and in the second case to 10 m/s (Figure \ref{mvehicle}).  
The former model represents walking speeds (e.g., soldiers) while
the later one represents vehicle speeds (e.g., tanks). 

  The results presented here are the average of runs over 3 random
  topologies. 
  The error calculation is done based on the distance
  between two nodes at the time of reading. 
  Note that for the static network simulations we previously
  discussed, 
  nodes chose a random number between 0 and 100 as their initial value
  and then during the reporting phase, each sensor incremented its
  reading by a fixed amount (10 each second) at fixed intervals.  In
  the revised data model, the variations in the sensor readings are
  not so high. Thus, the results with the revised data model and the
  initial data model are not comparable numerically per se.  However,
  one can clearly see the same trend in the relative performance of the
  different protocols. 
  Figures~(\ref{sp2f5} and \ref{sp2f2})
  and~(\ref{sp10f5}, \ref{sp10f2}, and \ref{sp10f1}) show that
  RFiltercast and both the randomized protocols perform better than
  or very close to that of flooding while the error value for 
  Filtercast is high. Similar to its performance in static networks, for low
  data rates, flooding starts performing better in terms of accuracy
  than all the other protocols.

\begin{figure*}
\begin{center}
\mbox{
\subfigure[speed 2 m/sec.\label{sp2werreng}]{\epsfig{file=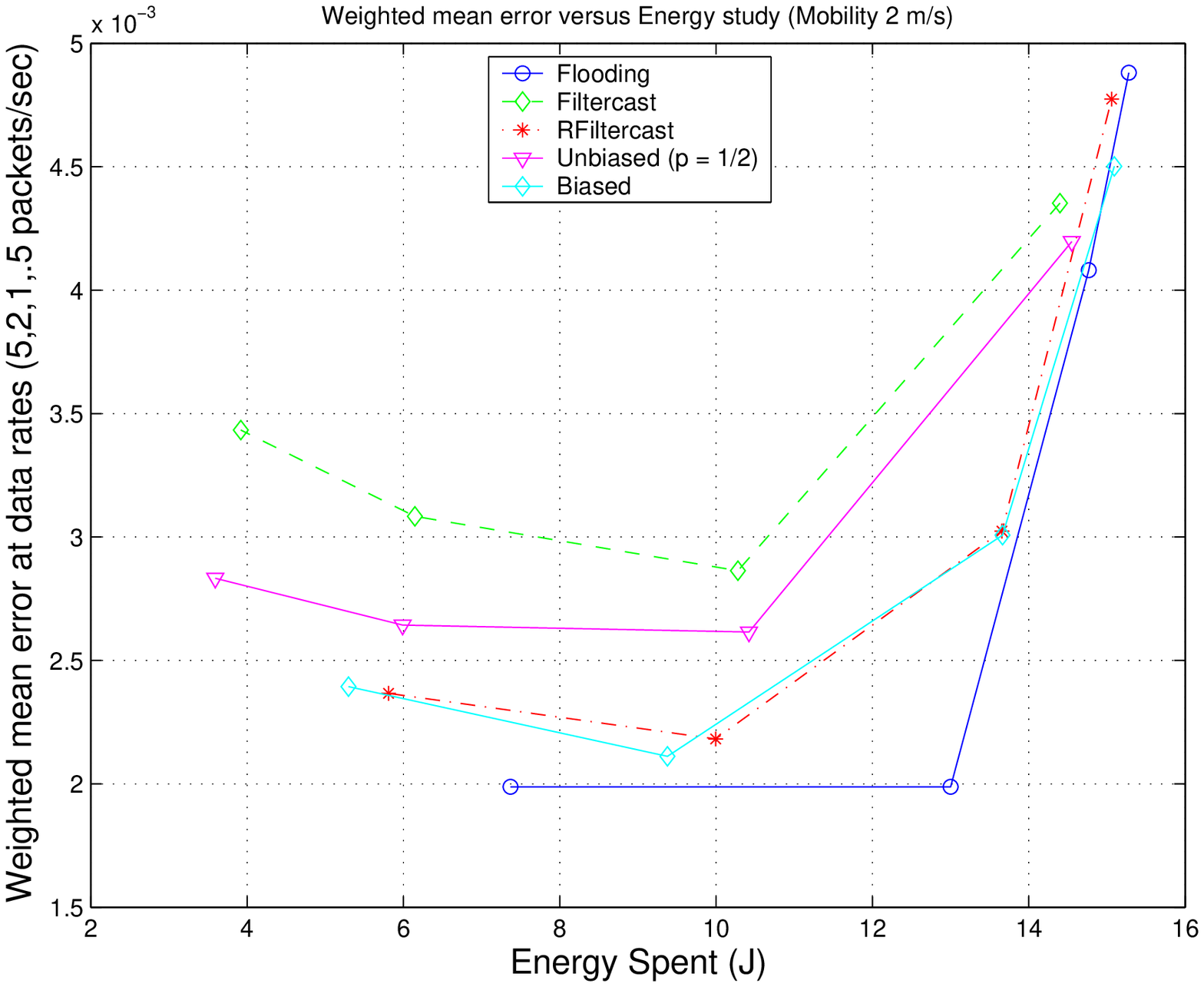,scale=0.4}} \quad
\subfigure[speed 10 m/sec.\label{sp10werreng}]{\epsfig{file=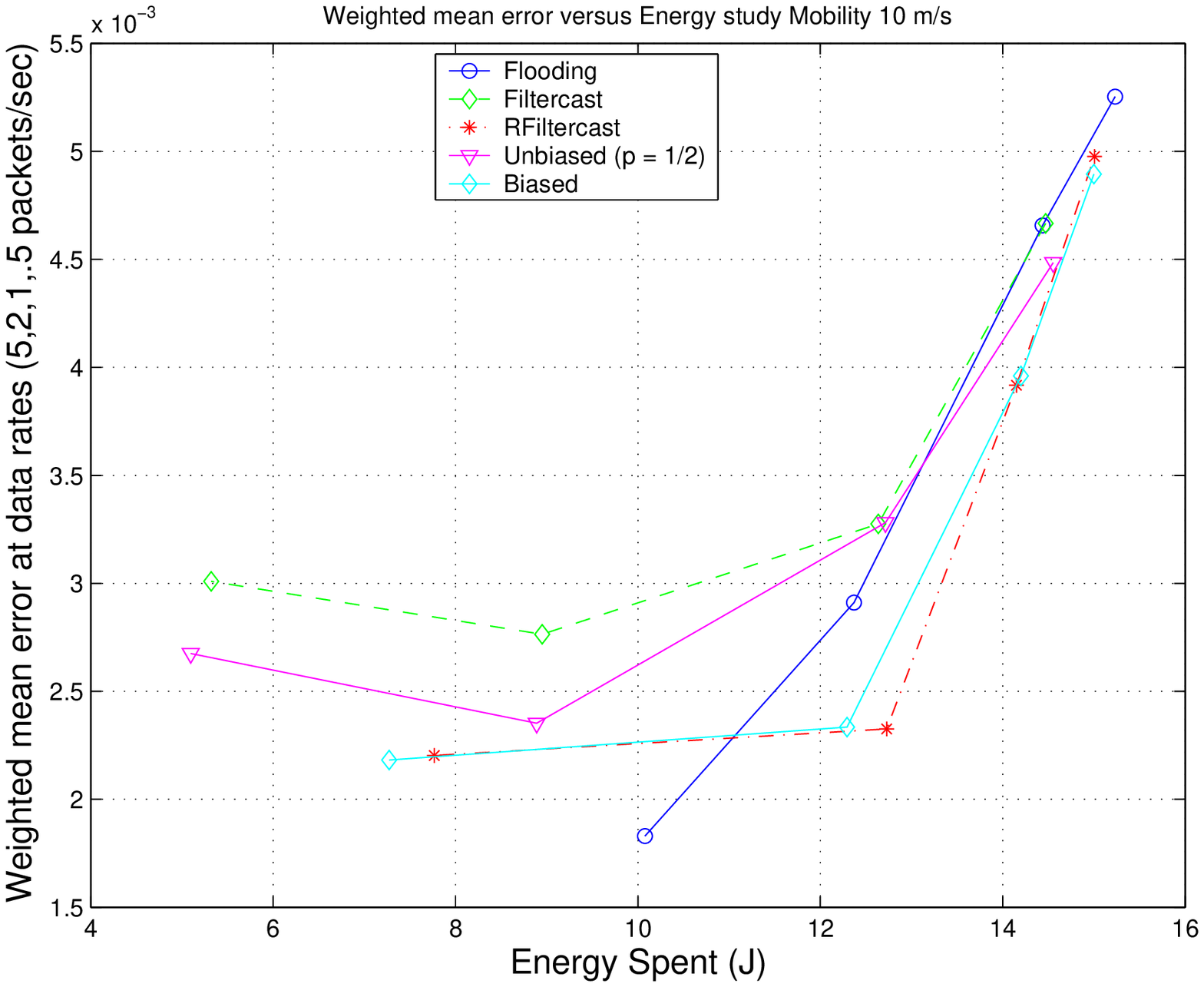,scale=0.4}}
}
\caption{Mobile sensors:  Weighted error-energy tradeoff.}
\label{mwerreng} 
\end{center}
\end{figure*}     

Figure \ref{mwerreng} shows the trade-off between
energy and weighted error, using the weighted error calculation method 
described in Eqns.~\ref{erroreqn1} and \ref{erroreqn2} for both mobility 
cases. In these figures, the X-axis indicates the energy spent in Joules and 
the Y-axis shows mean weighted error. The trends observed are very similar to 
that of the static network (Figure~\ref{werrenggridstatic}). At high data 
rates, 
flooding spends maximum energy and has the highest error. Note that
for higher mobility (Figure~\ref{sp10werreng}), the performance of flooding 
is worse than that of the low mobility case (Figure~\ref{sp2werreng}). In 
the case
of high mobility, the performance of RFiltercast and the biased randomized 
protocol is better than flooding except for the lowest data rate.
These results indicate that the protocols are resilient to
mobility, as changes in speed have almost no impact on the error values
of the protocols for a realistic data model.

\section{Discussion} \label{discussion}
Overall from these results, we can conclude the following: in the case
of applications that can exploit non-uniform information, protocols
can be designed to make efficient use of the available bandwidth while
providing the necessary level of accuracy.  Generally, RFiltercast
outperforms Filtercast when the network is not congested. Also, naive,
randomized protocols such as the unbiased protocol, outperform
specialized protocols such as Filtercast.  This is because in general
these protocols forward messages more aggressively with the parameters
we selected.  In our setting, since accuracy is a function of
distance, the errors for far sensors count less and thus overall these
protocols perform well.  The biased randomized protocol has comparable
performance to that of RFiltercast. 

In the simulations presented here, the biased protocol performs better in 
terms of accuracy than the unbiased protocol, even for distant sink nodes.  
This effect is simply due to the forwarding probability settings and, most 
importantly, the size of the simulated networks, which limits the maximum 
number of hops between the source and destination. As mentioned previously, in 
the biased protocol sensors transmit packets from nearby sensors with high 
probability, and this probability decreases linearly as a function of the 
number of hops between the source and the sensor transmitting the 
packet (see section~\ref{biasedprob}). However, in the case of the unbiased protocol, for 
any packet that needs to be forwarded, the forwarding probability is constant 
(0.5 in our simulations). In our simulations, the maximum number of hops is 
limited to six, since we could not run larger simulations due to computational 
resource constraints. Therefore, for the simulations presented here, the biased
protocol has higher forwarding probability than the unbiased protocol
for the first few hops (with respect to the source), which dominates the 
picture. We conjecture that if we increase the number of hops (to, say, 20), 
then for distant sinks, the unbiased protocol will perform
better than the biased protocol in terms of accuracy. 
                                                                               Note that both Filtercast and
RFiltercast have some overhead to maintain the source lists and the
count of how many packets a given source node has transmitted. On the
other hand, the randomized protocols do not require such state to be 
maintained. Also, the randomized protocols are resilient to
mobility.  We believe that randomized protocols with intelligent
adjustments of forwarding probabilities can be considered as the most
efficient alternative for non-uniform data dissemination.

\section{Related Work} ~\label{related-work}

Recently, sensor networks have drawn a considerable amount of attention from
the research community.
Most of this existing work focuses on two primary cases: (1) Sensors
send their data toward a central base station that has infinite
power and is responsible for all data processing, and no
\textit{in-network} processing is done.  (2) Sensors do some
in-network data processing such as data fusion and this high level
data is sent to the central base station.  A number of such approaches
have been proposed (e.g.,
~\cite{heinzelman-00,intanagonwiwat-00,lindsey-01}).
However, in our case, we do not assume the presence of any such
base station, and the sensors disseminate information among themselves
so that the user can connect to any of the sensors to extract network
information.  

Other studies considered specific sensor network applications and
their implication on protocol design.  Cerpa et al.~\cite{cerpa-01},
have considered habitat monitoring and have designed protocols to
match the application need.
Heinzelman et al.~\cite{heinzelman-99}, described adaptive protocols
for information dissemination.  In this work, to save energy, sensors
send out advertisements for data they have, and they only send the
actual data if it is requested by one or more nodes.
In previous work~\cite{tilak-ms}, we described probabilistic flooding
alternatives.  However, the main goal of that work was congestion
avoidance rather than non-uniform information dissemination.  

Li et al.~\cite{li-02} proposed a gossip-based approach for routing protocols
to reduce routing overhead. However, the study focused only on routing 
messages (with implicit uniform information granularity requirement).
Recently, Barette et al.~\cite{barrette-03} proposed a family of 
gossip-based routing protocols for sensor networks. In their study 
they considered 
various parameters such as the number of hops between the source and the 
destination, the number of hops the packet has traveled, etc. 

In the DREAM~\cite{dream} routing protocol, 
routing tables are updated based on the distance between two nodes 
and the mobility rate of a given node.  While this work has a similar 
flavor to our work, exploiting non-uniform information needs, it is limited 
to only adjusting routing tables and does not apply to the actual data that is
exchanged between two nodes.

Kempe et al.~\cite{kempe-01} presented theoretical results for gossiping 
protocols with resource location as a motivating problem and delay 
as the primary consideration. 
In their setting, a node at distance $d$ from the origin of a new information 
source
should learn about it with a delay that grows slowly with $d$ and independent 
of network size. They do not consider application level performance 
criteria such as accuracy, which is part of our study.

In this paper we have considered flooding as one of the alternatives
for data dissemination in sensor networks. However, flooding and its
alternatives have also been explored in the context of mobile ad hoc
networks.  Perkins et al.  describe IP Flooding in ad-hoc
networks~\cite{FLDDRFT}.  While this paper considers probabilistic
flooding protocols for sensor networks, Sasson et al.~have studied
probabilistic flooding for ad hoc networks~\cite{probflood} and used
the phase transition phenomenon as a basis to select the broadcasting
probability.
Williams et al.~\cite{williams-02} described and compared several broadcasting 
protocols (including probabilistic protocols) in the context of
mobile ad hoc networks.

The primary difference between our work and existing work is the
application requirement. In our study, we focus on a new application
requirement, non-uniform information dissemination, and 
we analyze protocols for this class of applications.

\section{Conclusions and Future Work} \label{conclusion}

In this paper we considered sensor network applications where events
need to be disseminated to observers that may be present anywhere in
the sensor field.  For such applications, simply flooding all the data
is extremely wasteful.  Therefore, we defined the idea of non-uniform
information dissemination to capitalize on the fact that the value of data
is typically highest for observers that are closest to the source of the data.
We developed and analyzed several protocols to accomplish non-uniform
dissemination, both deterministic (Filtercast and
RFiltercast) and non-deterministic (unbiased and biased) protocols, and
we evaluated them under various traffic loads and transmission ranges and
with or without mobility.  In all cases, the developed protocols were
clearly superior to simple flooding, both from an application and a network
perspective.  With flooding, congestion appears to be a limiting
constraint and further, flooding is not generally
energy-efficient. Our results indicate that the performance of
RFiltercast and the biased randomized protocol is almost
equivalent. RFiltercast requires each sensor to maintain some extra
state information, whereas the biased randomized protocol is
completely stateless and has negligible overhead. Also, we note that
the performance of Filtercast and the unbiased randomized protocol is
almost equivalent.  We also showed that RFiltercast as well as the
randomized protocols are resilient to mobility.

While in this paper the distance between two nodes is used as a parameter for
non-uniform data dissemination,
in our future work
we will focus on a broad range of applications with a non-uniform information 
dissemination requirement, where factors other than distance, such as 
importance of the information and confidence in the
generated data, can be used. Also, we would like to develop protocols that will 
tune the forwarding probabilities dynamically depending upon factors such as
traffic load, network connectivity, resources (remaining battery power), etc. We believe 
that this will be an important step towards making these networks self-configuring.

We would also like to develop a priority-based protocol where a source 
marks all its outgoing packets with a certain priority to indicate the importance of
the information contained in the given packet. Any forwarding node can consider
the priority of the packet when making its forwarding decision. 
These techniques will
extend the applicability of non-uniform information dissemination to new
classes of applications for wireless sensor networks.

\bibliography{sensor-nets}

\begin{thebibliography}{10}

\bibitem{barrette-03}
C.~L. Barrett, S.~J. Eidenbenz, L.~Kroc, M.~Marathe, and J.~P. Smith.
\newblock Parametric probabilistic sensor network routing.
\newblock In {\em Proceedings of the 2nd ACM international conference on
  Wireless sensor networks and applications}, pages 122--131. ACM Press, 2003.

\bibitem{cerpa-01}
A.~Cerpa, J.~Elson, D.~Estrin, L.~Girod, M.~Hamilton, and J.~Zhao.
\newblock Habitat monitoring: Application driver for wireless communications
  technology.
\newblock In {\em Proc. ACM SIGCOMM Workshop on Data Communications in Latin
  America and the Caribbean}, Apr. 2001.

\bibitem{kempe-01}
A.~D. David~Kempe, Jon~Kleinberg.
\newblock Spatial gossip and resource location protocols.
\newblock In {\em Annual ACM Symposium on Theory of Computing (STOC)}, 2001.

\bibitem{heinzelman-00}
W.~Heinzelman, A.~Chandrakasan, and H.~Balakrishnan.
\newblock Energy-efficient routing protocols for wireless microsensor networks.
\newblock In {\em Proc. 33rd Hawaii International Conference on System Sciences
  (HICSS '00)}, Jan. 2000.

\bibitem{heinzelman-99}
W.~Heinzelman, J.~Kulik, and H.~Balakrishnan.
\newblock {Adaptive Protocols for Information Dissemination in Wireless Sensor
  Networks}.
\newblock In {\em {Proceedings of the Fifth Annual ACM/IEEE International
  Conference on Mobile Computing and Networking (MobiCom '99)}}, pages
  174--185, Aug. 1999.

\bibitem{FLDDRFT}
{IETF MANET Working Group Internet Draft-- IP Flooding in Ad hoc Mobile
  Networks draft-ietf-manet-bcast-00.txt}.
\newblock
  {http://www.ietf.org/proceedings/01dec/I-D/draft-ietf-manet-bcast-00.txt},
  2001.

\bibitem{intanagonwiwat-00}
C.~Intanagonwiwat, R.~Govindan, and D.~Estrin.
\newblock Directed diffusion: A scalable and robust communication paradigm for
  sensor networks.
\newblock In {\em Proc. 6th ACM International Conference on Mobile Computing
  and Networking (Mobicom'00)}, Aug. 2000.

\bibitem{li-02}
L.~Li, J.~Halpern, and Z.~Haas.
\newblock Gossip-based ad hoc routing.
\newblock In {\em IEEE Infocom}, 2002.

\bibitem{lindsey-01}
S.~Lindsey, C.~Raghavendra, and K.~Sivalingam.
\newblock Data gathering in sensor networks using energey-delay metric.
\newblock In {\em Proc. International Workshop on Parallel and Distributed
  Computing Issues in Wireless Networks and Mobile Computing}, Apr. 2001.

\bibitem{ns-2}
{Network Simulator}.
\newblock {http://isi.edu/nsnam/ns}.

\bibitem{dream}
B.~S., I.~Chlamtac, V.~R. Syrotiuk, and B.~A. Woodward.
\newblock A distance routing effect algorithm for mobility (dream).
\newblock In {\em Fourth Annual ACM/IEEE International Conference on Mobile
  Computing and Networking}, 1998.

\bibitem{magnetos}
E.~G. Sirer, R.~Barr, T.~D. Kim, and I.~Y.~Y. Fung.
\newblock Automatic code placement alternatives for ad hoc and sensor
  networks., 2001.

\bibitem{tilak-ms}
S.~Tilak.
\newblock The role of the infrastructure in sensor networks, 2002.
\newblock M.S. thesis- Computer Science department, SUNY Binghamton.

\bibitem{tilak-wsna}
S.~Tilak, N.~Abu-Ghazaleh, and W.~Heinzelman.
\newblock Infrastructure tradeoffs for sensor networks.
\newblock In {\em First ACM International Workshop on Wireless Sensor Networks
  and Applications}, 2002.

\bibitem{tilak-icnp03}
S.~Tilak, A.~Murphy, and W.~Heinzelman.
\newblock Non-uniform information dissemination for sensor networks.
\newblock In {\em The 11th IEEE International Conference on Network Protocols
  (ICNP'03)}, Nov. 2003.

\bibitem{williams-02}
B.~Williams and T.~Camp.
\newblock Comparison of broadcasting techniques for mobile ad hoc networks.
\newblock In {\em Proceedings of the ACM International Symposium on Mobile Ad
  Hoc Networking and Computing (MOBIHOC)}, 2002.

\bibitem{probflood}
A.~e.~S. Yoav~Sasson, David~Cavin.
\newblock Probabilistic broadcast for flooding in wireless mobile ad hoc
  networks, 2002.

\end{thebibliography}
\bibliographystyle{ieee}
\end{document}